\documentclass[prl,aps,twocolumn,showpacs]{revtex4}
 \newcommand{\mytitle}[1]{
 \twocolumn[\hsize\textwidth\columnwidth\hsize
 \csname@twocolumnfalse\endcsname #1 \vspace{1mm}]}
 \newcommand{\beq}{\begin{equation}}
 \newcommand{\eeq}{\end{equation}}
 \newcommand{\bea}{\begin{eqnarray}}
 \newcommand{\eea}{\end{eqnarray}}

\usepackage{graphics}
\usepackage{graphicx}
\usepackage{amssymb}
\usepackage{bm}

\begin{document}

\title{Dynamical properties of a nonequilibrium quantum dot 
close to localized-delocalized quantum phase transitions}
\author{Chung-Hou Chung$^{1,2}$}
\affiliation{
$^{1}$Electrophysics Department, National Chiao-Tung University, HsinChu, Taiwan R.O.C. 300\\
$^{2}$Departments of Physics and Applied Physics, Yale University, New Haven, CT, 06511 USA\\
}

\date{\today}

\begin{abstract}
We calculate the dynamical decoherence rate and susceptibility 
of a nonequilibrium quantum dot close to the  
delocalized-to-localized 
quantum phase transitions. The setup concerns   
a resonance-level coupled to two spinless fermionic 
baths with a finite bias voltage and an Ohmic bosonic bath representing 
the dissipative environment. 
 The system is equivalent to an anisotropic Kondo model. 
 As the dissipation  
strength increases, the system at zero temperature and zero bias 
show quantum phase transition  
between a conducting delocalized phase to an insulating localized 
phase. Within the nonequilibrium functional Renormalization Group (FRG) 
approach, we address the finite bias crossover in dynamical decoherence 
rate and charge susceptibility close to the phase transition. 
We find the dynamical decoherence rate increases with increasing 
frequency. In the delocalized phase, 
it shows a singularity  
at frequencies equal to positive or negative 
bias voltage.  
As the system crossovers to 
the localized phase, the decoherence rate at low frequencies 
get progressively 
smaller and this sharp feature is gradually smeared out, 
leading to a single linear frequency dependence. 
The dynamical charge susceptibility shows a dip-to-peak 
crossover across the delocalized-to-localized transition. 
Relevance 
of our results to the experiments is discussed.

\end{abstract}
\maketitle

\subsection{\bf Introduction}

Quantum phase transitions (QPTs)\cite{sachdevQPT,Steve} 
due to competing quantum ground states 
are of fundamental importance in condensed matter physics 
and have attracted much attention both theoretically and 
experimentally. 
Near the transitions, exotic quantum critical properties are 
realized. In recent years, there has been 
a growing interest in QPTs in nanosystems\cite{lehur1,lehur2,
zarand,Markus,matveev,Zarand2}. 
Very recently, QPTs have been extended to nonequilibrium 
nanosystems with a large bias voltage being applied to the setups. 
Close to QPTs  
much of the attention has been 
focused on equilibrium properties; while relatively less is known 
on the nonequilibrium properties. The key difference between 
equilibrium and nonequilibrium properties near QPTs is the voltage-induced 
nonequilibrium decoherence rate which behaves very differently 
from that in equilibrium 
at finite temperatures, 
leading to distinct nonequilibrium properties near QPTs.

Recently, two generic 
examples\cite{chung} have been studied:
(i). the transport through a dissipative 
resonance-level (spinless quantum dot)  
at a finite bias voltage where dissipative bosonic 
bath (noise) coming from the environment in the leads\cite{chung}, 
(ii). a spinful quantum dot coupled to two interacting Luttinger 
liquid leads\cite{chung2} where the electron interactions can be regarded as 
an effective Ohmic dissipative bosonic bath\cite{Florens}.  
 As dissipation (or interaction) strength is increased, both systems 
can be mapped onto different effective 
 Kondo models, which exhibit QPT in transport 
from a conducting (delocalized) phase where resonant tunneling 
dominates and an insulating (localized) phase where 
the dissipation (or electron-electron interaction) prevails. 
Similar dissipation driven QPTs have been investigated 
in various systems\cite{Josephson,McKenzie}.  To obtain the nonequilibrium 
transport properties, the authors 
in Ref.\cite{chung} and Ref.~\cite{chung2} applied the  
nonequilibrium Renormalization Group (RG) approach\cite{Rosch} 
in the form of self-consistent scaling equations for frequency-dependent Kondo 
couplings and the static decoherence rate $\Gamma(V,T,B)$. 
Though the dynamical nonequilibrium effects in Kondo models  
have been addressed\cite{schoeller,kehrein,coleman}, less is known 
of the steady-state nonequilibrium decoherence effect on the 
anisotropic and/or two-channel Kondo models.  
In this paper, we address the nonequilibrium decoherence effect on the 
anisotropic Kondo model in the presence of a large 
bias voltage, which is relevant for the 
delocalized-to-localized nonequilibrium QPTs 
in the dissipative quantum dot (the first example mentioned above). 
To obtain the dynamics (frequency dependence) 
of decoherence rate, we generalize the approach taken 
in Ref.~\cite{chung,chung2} via a Functional Renormalization 
Group (FRG) method developed in Ref.~\cite{woelfleFRG}. 
The nonequilibrium decoherence rate is directly proportional to 
the width of the peak in dynamical spin susceptibility.    
We furthermore investigate  
the spectral properties of the 
dynamical decoherence rate close to the QPT and its implications 
to the dynamical charge susceptibility, which can be measured 
experimentally.  
In particular, as the system 
goes from the delocalized to the localized phase we find 
the dynamical decoherence rate for small frequencies 
gets smaller in magnitude  
and the singular ``kink-like'' behavior occurring 
at the frequencies equal to the bias voltage ($\omega=\pm V$) 
gets smeared out. We have calculated the dynamical spin susceptibility. 
As the system moves from the delocalized to the localized phase,   
it shows a dip-to-peak crossover and the smearing of  
the sharp feature at $\omega\approx \pm V$. The relevance 
of our results to the experiments is discussed.

\subsection{\bf   Dissipative resonant level model}

{\bf \it Model Hamiltonian.} The system we study here is a  
spin-polarized quantum dot coupled to two Fermi-liquid 
leads subjected to noisy Ohmic environment, which coupled capacitively to 
the quantum dot\cite{chung}. 
%The noisy environment here 
%consists of a collection of harmonic oscillators with the Ohmic correlation:
%$G_{\phi}({\it i}\omega) \equiv <\phi({\it i}\omega) \phi(-{\it i} \omega)> = 
%2\pi \frac{R}{R_k} [|\omega|+\frac{\omega^2}{\omega_c}]^{-1}$ with 
%$R$ being the circuit resistance and $R_k\equiv 2\pi \hbar/e^2\approx 25.8k\Omega$ 
%being the quantum resistance. 
For a dissipative resonant level (spinless quantum dot) 
model, the quantum phase transition  
separating the conducting and insulating phase for the level is solely 
driven by dissipation. 
The Hamiltonian is given by\cite{chung}:
\begin{eqnarray}
H &=& \sum_{k,i=1,2} (\epsilon(k)-\mu_i) c^{\dagger}_{k i}c_{k i} + t_{i}
c^{\dagger}_{ki} d + h.c.  \nonumber \\
&+& \sum_{r} \lambda_{r} (d^{\dagger}d-1/2) (b_{r} + b^{\dagger}_{r}) +
\sum_{r} \omega_{r} b^{\dagger}_{r} b_{r},\nonumber\\
&+& h (d^{\dagger} d -1/2)
\end{eqnarray}
where $t_{i}$ is the hopping amplitude between the
lead $i$ and the quantum dot, $c_{ki}$ and $d$ are electron operators for the 
Fermi-liquid leads and the quantum dot, respectively, 
$\mu_i = \pm V/2$ is the chemical potential (bias voltage) 
applied on the lead $i$, while $h$ is the energy level of the dot.
We assume
that the electron spins have been polarized by a 
strong magnetic field. Here, $b_{\alpha}$ are the boson operators of the
dissipative bath with an ohmic spectral density
\cite{lehur2}: $\mathit{J}(\omega) = \sum_{r} \lambda_{r}^2
\delta(\omega-\omega_{r}) = \alpha \omega$ with $\alpha$ being the 
strength of the dissipative boson bath.\\ 

\begin{figure}[t]
\begin{center}
\includegraphics[width=8cm]{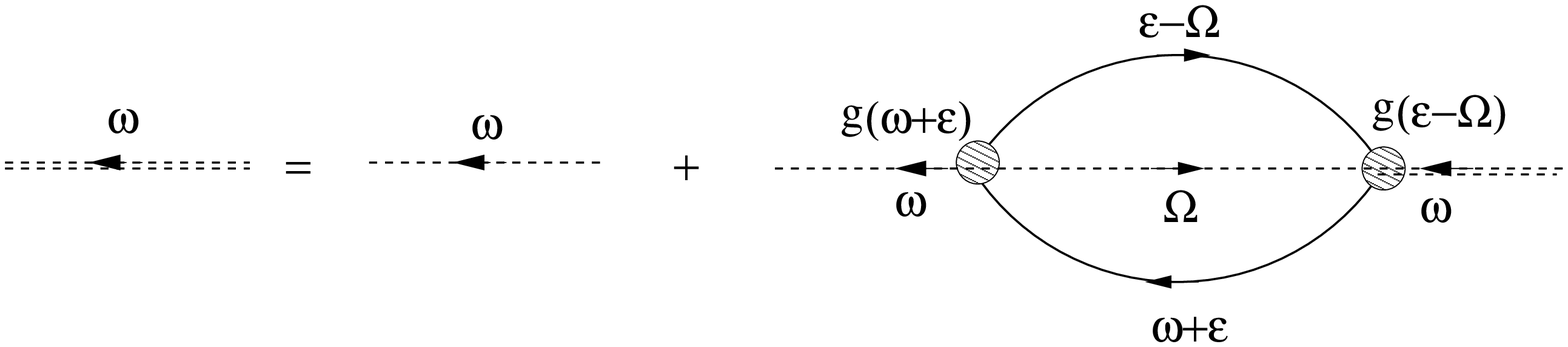}
\end{center}
\par
\vspace{-0.5cm} 
\caption{(Color online) Diagram for the pseudofermion self energy.}
\label{self}
\end{figure}

Through similar bosonization and
refermionization procedures as in equilibrium 
\cite{lehur1,lehur2,Markus,matveev}, 
the above model is maped onto 
an equivalent anisotropic Kondo model in an effective 
magnetic field $h$ with the effective left $L$ and right $R$ Fermi-liquid 
leads\cite{chung}. The effective Kondo 
model takes the form:
\begin{eqnarray}
{H}_{K} &=&\sum_{k,\gamma =L,R,\sigma =\uparrow ,\downarrow }[\epsilon
_{k}-\mu _{\gamma }]c_{k\gamma \sigma }^{\dagger }c_{k\gamma \sigma } \nonumber \\
&+&(J_{\perp }^{1}s_{LR}^{+}S^{-}+J_{\perp
}^{2}s_{RL}^{+}S^{-}+h.c.)\nonumber \\
&+&\sum_{\gamma =L,R}J_{z}s_{\gamma \gamma
}^{z}S^{z} + h S_z,
\label{Hk}
\end{eqnarray}
where $c_{kL(R)\sigma }^{\dagger }$ is the electron operator of the
effective lead $L(R)$, with spin $\sigma$. 
Here, the spin operators are related to the electron operators on the dot 
by: $S^{+}=d^{\dagger }$, $S^{-}=d$, and $S^{z}=d^{\dagger}d -1/2 = n_d -1/2$ 
where $n_d=d^{\dagger }d$ describes the charge occupancy of the level.
The spin operators for electrons in the effective leads are 
$s_{\gamma \beta }^{\pm }=\sum_{\alpha ,\delta ,k,k^{\prime }}1/2c_{k\gamma
\alpha }^{\dagger }\mathbf{\sigma }_{\alpha \delta }^{\pm }c_{k^{\prime
}\beta \delta }$, 
the transverse and longitudinal Kondo couplings are given by 
$J_{\perp }^{1(2)}\propto {t_{1(2)}}$ and 
$J_{z}\propto 1/2(1-{1}/\sqrt{2\alpha ^{\ast }})$ respectively, 
and the effective bias voltage is 
$\mu _{\gamma }=\pm \frac{V}{2}\sqrt{1/(2\alpha ^{\ast })}$, where $%
1/\alpha ^{\ast }=1+\alpha $. Note that $\mu _{\gamma }\rightarrow \pm V/2$
near the transition ($\alpha ^{\ast }\rightarrow 1/2$ or 
$\alpha \rightarrow 1$) where the above mapping is exact. 
The spin operator of the quantum dot in the effective Kondo model 
$\vec{S}$ can also 
be expressed in terms of spinful pseudofermion operator 
$f_{\sigma}$: 
$S_{i=x,y,z} = f^{\dagger}_{\alpha} \sigma_{i=x,y,z}^{\alpha\beta} f_{\beta}$. 
In the Kondo limit where only the singly occupied fermion states are physically 
relevant, a projection onto the singly occupied states is necessary in the 
pseudofermion representation\cite{Rosch,chung}.
This can be achieved by introducing 
the Lagrange multiplier $\lambda$ so that 
$Q=\sum_{\gamma} f^{\dagger}_{\gamma} f_{\gamma}=1$\cite{paaske,latha}. 
%An observable $\mathcal{A}$ is defined as\cite{Rosch}:
%\begin{equation}
%<\mathcal{A}>_{Q=1} =lim_{\lambda\to \infty} 
%\frac{<\mathcal{A} Q>_{\lambda}}{<Q>_{\lambda}}
%\end{equation}
%Note that the above mapping 
%can be generalized to a resonance-level coupled to spinless Luttinger 
%liquid leads (see Appendix), which has relevance for  
%edge state tunneling in Quantum Hall states. 
In equilibrium, the above anisotropic Kondo model 
exhibits the Kosterlitz-Thouless (KT) transition from a delocalized 
phase with a finite conductance $G\approx \frac{1}{2\pi\hbar}$ 
($e=\hbar=1$) 
for $J_{\perp}+ J_z > 0$ to a localized phase for 
$J_{\perp}+ J_z \le 0$ with vanishing conductance. The distinct profile 
in nonequilibrium transport near the localized-delocalized 
KT transition 
has been addressed in Ref. \cite{chung}. Below we will turn our attention 
to the dynamical charge susceptibility of the quantum dot close to 
the transition.\\

{\bf \it Nonequilibrium FRG formalism.} 
The non-equilibrium perturbative FRG 
approach is based on the generalization of the perturbative 
RG approach studied in Ref.~\cite{Rosch} for the nonequilibrium 
Kondo model. Following Ref.~\cite{woelfleFRG}, the 
frequency dependent RG scaling equations for the effective 
Kondo couplings in the Keldysh 
formulation are given by\cite{Rosch}:
\begin{eqnarray}
\frac{\partial g_{z}(\omega )}{\partial \ln D} &=&-\sum_{\beta =-1,1}\left[
g_{\perp }\left( \frac{\beta V}{2}\right) \right] ^{2}\Theta _{\omega +
\frac{\beta V}{2}}  \label{gpergz} \nonumber \\
\frac{\partial g_{\perp }(\omega )}{\partial \ln D} &=&-\sum_{\beta=-1,1}  g_{\perp }\left( \frac{\beta V}{2}\right)
 g_{z}\left( \frac{\beta V}{2}\right) \Theta _{\omega +\frac{\beta V}{2}},
\label{RG-Kondo}
\end{eqnarray}
where $g_{\perp}(\omega) = N(0)J_{\perp}^{1} = N(0)J_{\perp\sigma}^{2}$, 
$g_{z}(\omega) = N(0) J_{z}$ are dimensionless 
frequency-dependent Kondo couplings with $N(0)$ being 
density of states per spin of the 
conduction electrons (we assume symmetric hopping 
$t_1=t_2=t$). Here, 
$\Theta _{\omega }=\Theta (D-|\omega +\mathit{i}\Gamma(\omega) |)$ (with 
$D<D_{0}$ being the running cutoff) comes from the leading
logarithmic corrections for the Kondo vertex function originated 
from the product of the Keldysh component of the lead electron Green function 
$G_{\alpha}^{K}(\omega)$ with the real part of the retarded or advanced 
dressed pseudo fermion propagator $Re(\tilde{G}_f^{R/A})$\cite{woelfleFRG}:
\begin{eqnarray}
G_{\alpha}^K(\omega) &=& -2\pi {\it i} \tanh(\frac{\omega-\mu_{\alpha}}{2T}) N_0 
\Theta(D_0-|\omega|)\nonumber \\
\tilde{G}_{f\sigma}^R(\omega) &=& \frac{1}{\omega- \Sigma_{\sigma}^R(\omega)} = 
[\tilde{G}_{f\sigma}^A(\omega)]^{\ast}, 
\end{eqnarray}
where $\Sigma_{\sigma}^R(\omega)= Re(\Sigma_{\sigma}^R(\omega)) + 
\frac{{\it i}}{2} \Gamma_{\sigma}(\omega)$ is the impurity self-energy 
(defined below) with the imaginary part being the dynamical 
decoherence rate  $\Gamma_{\sigma}(\omega)$, and $N_0=\frac{1}{2D_0}$. 
At $T=0$, the above correction 
can be approximated by the $\Theta$-function in the RG equations shown above. 
Also, the dynamical decoherence (dephasing) rate $\Gamma_{\sigma}(\omega)$ 
at finite bias which serves as a cutoff to 
the RG flow of $g_{\perp,z}(\omega)$\cite{Rosch}, and it is 
determined self-consistently along with the RG equations 
for the Kondo couplings. Note that in general $\Gamma_{\sigma}(\omega)$ 
depends on the impurity spin $\sigma$; however, in the absence of 
the magnetic field as we consider here, we have the spin symmetry and hence: 
$\Gamma_{\uparrow}(\omega)=\Gamma_{\downarrow}(\omega)=\Gamma(\omega)$. 
We can obtain $\Gamma(\omega)$    
by the imaginary part of the pseudofermion self energy via second-order 
renormalized perturbation theory (see Fig.~\ref{self}):

\begin{figure}[t]
\begin{center}
%\vspace{0.1cm}
%\epsfig{file=gpergz.eps,width=0.9\linewidth}
\includegraphics[width=8.5cm]{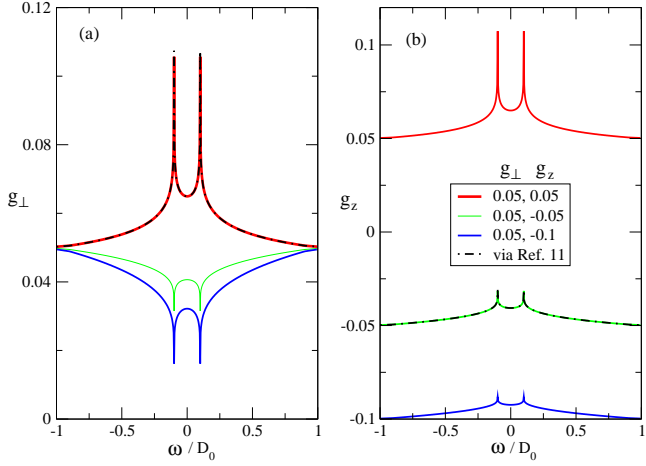}
\end{center}
\par
\vspace{-0.5cm} %\label{figgpergz}
\caption{ Frequency dependent Kondo couplings  
of a dissipative resonant level 
model at zero temperature for (a). $g_{\perp}(\protect\omega)$ and (b). 
$g_{z}(\protect\omega)$  across the localized-delocalized 
transition for different 
bare Kondo couplings via the FRG approach (solid lines) and the approach 
in Ref.~\cite{chung} (dot-dashed lines). We have set
$V = 0.2 D_0$ where $D_0 = 1$ for all the figures. }
\vspace{0.3cm}
\label{gpgzfig}
\end{figure}

\begin{eqnarray}
\Gamma_{\sigma}(\omega) &=& Im(\Sigma_{\sigma}(\omega)) = {\it i} 
(\Sigma_{\sigma}^R - \Sigma_{\sigma}^A)
\nonumber \\
{\it i} \Sigma_{\sigma}^{R(A)}(\omega) &=& \sum_{\alpha,\beta=L,R}
\frac{\theta_{\sigma\sigma'}}{16}\int 
\frac{d\epsilon}{2\pi} g_{\alpha\beta}(\omega+\epsilon) \nonumber \\
& & \chi_{cf,\sigma'}^{<(>),\alpha\beta}(\epsilon) 
G_{\beta}^{>(<)}(\epsilon+\omega)\nonumber \\
\label{Gamma-omega}
\end{eqnarray}
where $\sigma=\uparrow,\downarrow$, 
$\theta_{\sigma\sigma'}$ is the tensor associated with 
the product of the Pauli matrices\cite{Rosch}:
\begin{equation}
\theta_{\gamma\gamma'} = \frac{1}{2}\sum_{\sigma,\sigma'} 
\tau^i_{\sigma\sigma'}\tau^i_{\gamma\gamma'}
\tau^j_{\sigma'\sigma}\tau^j_{\gamma'\gamma} = \delta_{\gamma\gamma'} + 
2 \tau^1_{\gamma'\gamma}
\end{equation}
, and $\chi_{cf,\sigma}^{<(>),\alpha\beta}$ reads:
\begin{equation}
\chi_{cf,\sigma}^{<(>),\alpha\beta}(\epsilon) = 
\int \frac{d\Omega}{2\pi} g_{\beta\alpha}(\epsilon-\Omega) 
(\hat{G}_{\alpha\sigma}(\epsilon+\Omega) \hat{G}_{f\sigma}(\Omega))^{<(>)},
\end{equation}
where $\hat{G}$ is the Green's function in $2\times 2$ Keldysh space, and its 
lesser and greater Green's function 
are related to its retarded, advanced, and Keldysh components by:
\begin{eqnarray}
G^< &=& (G^K - G^R + G^A)/2\nonumber \\
G^> &=& (G^K +G^R - G^A)/2
\end{eqnarray}
Specifically, the lesser and greater components of Green's function 
of the conduction electron in the leads and of the quantum dot (impurity) 
are given by (in the absence of magnetic field):
\begin{eqnarray}
G_{L/R}^<(\epsilon) &=& {\it i} A_c(\epsilon) f_{\epsilon-\mu_{L/R}}\nonumber \\
G_{L/R}^>(\epsilon) &=& {\it i} A_c(\epsilon) (1-f_{\epsilon-\mu_{L/R}})\nonumber \\
G_{f\sigma}^<(\epsilon) &=& 2\pi {\it i} \delta(\epsilon) n_{f\sigma}(\epsilon)\nonumber \\
G_{f\sigma}^>(\epsilon) &=& 2\pi {\it i} \delta(\epsilon) (n_{f\sigma}(\epsilon) -1)
\end{eqnarray}
where $A_c(\epsilon)= 2\pi N_0^2 \Theta(D_0- \epsilon)$ is the density of states of the leads, 
$n_{f\sigma}(\epsilon)=f^{\dagger}_{\sigma}f_{\sigma}$ is 
the occupation number of the pseudofermion which 
obeys $n_{f\uparrow}+n_{f\downarrow}=1$, 
$n_{f\sigma}(\epsilon\rightarrow 0)= 1/2$ in the delocalized phase and 
$n_{f\uparrow}(\epsilon\rightarrow 0)\rightarrow 0$, 
$n_{f\downarrow}(\epsilon\rightarrow 0)\rightarrow 1$ in the localized 
phase\cite{chung,latha}. Here, the pseudofermion occupation number 
$n_{f\sigma}$ and the occupation number on the dot 
$n_d$ are related via $<n_{f\uparrow}-n_{f\downarrow}>= <n_d>-1/2$. 
Also, $f_{\omega-\mu_{L/R}} $ is  
the Fermi function of the $L/R$ lead 
given by $f_{\omega-\mu_{L/R}} = 1/(1+e^{(\omega-\mu_{L/R})/k_B T}))$.  
We have therefore:
\begin{eqnarray}
\chi_{cf,\sigma}^{<,\alpha\beta}(\epsilon) &=& 2\pi {\it i} 
g_{\beta\alpha}(\epsilon) A_c(\epsilon) (1-n_{f\sigma}(\epsilon)) 
f_{\epsilon-\mu_\alpha}\nonumber \\
\chi_{cf,\sigma}^{>,\alpha\beta}(\epsilon) &=& 2\pi {\it i} 
g_{\beta\alpha}(\epsilon) A_c(\epsilon) n_{f\sigma}(\epsilon) 
(1-f_{\epsilon-\mu_\alpha})
\end{eqnarray}

\begin{eqnarray}
\Gamma(\omega) =  
\frac{3}{4\pi} &\int{d\epsilon}& g_{\perp}(\epsilon+\omega) 
g_{\perp}(\epsilon) 
[f_{\epsilon-\mu_L}(1-f_{\epsilon+\omega-\mu_R})]\nonumber \\
&+ & g_{z}(\epsilon+\omega) 
g_{z}(\epsilon) 
[f_{\epsilon-\mu_L}(1-f_{\epsilon+\omega-\mu_L})]\nonumber \\
&+& (L\rightarrow R)
\label{gamma}
\end{eqnarray}
The FRG approach here is accomplished by self-consistently 
solving the RG scaling equation Eq.~\ref{RG-Kondo}  
subject to Eq.~\ref{Gamma-omega} and Eq.~\ref{gamma}. 
The solutions at zero temperature 
for $g_{\perp}(\omega)$ and $g_{\sigma, z}(\omega)$ across the transition 
are shown in Fig. \ref{gpgzfig}\cite{chung}. Note that our FRG approach 
is somewhat different from that in Ref.~\cite{woelfleFRG}: 
We do not formulate and solve for the RG scaling equation for the 
impurity self energy $\Sigma_{\sigma}(\omega)$ as shown in 
Ref.~\cite{woelfleFRG}; instead, we calculate the imaginary part 
of the impurity self-energy (or the decoherence rate $\Gamma(\omega)$) 
self-consistently via second-order renormalized perturbation theory. 
Nevertheless, we have checked in the simple Kondo limit with isotropic 
couplings that the frequency-dependent renormalized Kondo coupings 
$g_{\perp,z}(\omega)$ obtained here agree very well with 
that obtained via an equivalent FRG approach in Ref.~\cite{woelfleFRG}).\\ 
 
As the system goes from the delocalized to localized phase transition, 
the features in $g_{\perp}(\omega)$ at $\omega=\pm V/2$ 
undergoes a crossover from symmetric double peaks to symmetric double dips, 
while the symmetric two peaks in $g_{z}(\omega=\pm V/2)$ still 
remain peaks. We find the above results based on the more rigorous FRG approach 
are in good agreement with the previous heuristic method\cite{chung}, 
which provides us with an independent and consistency check 
on the previous results. 
Note, however that from previous 
approach in Ref.~\cite{chung} and \cite{chung2} 
the decoherence rate was taken approximately as 
$\Gamma(\omega=0)$; we now generalize this $\Gamma$ by including   
the frequency dependence. This generalization 
improves the previous RG formalism and 
it also provides us with 
more features in the dynamical quantities across the transition, 
such as in dynamical charge susceptibility. 
\begin{figure}[t]
\begin{center}
\includegraphics[width=8.5cm]{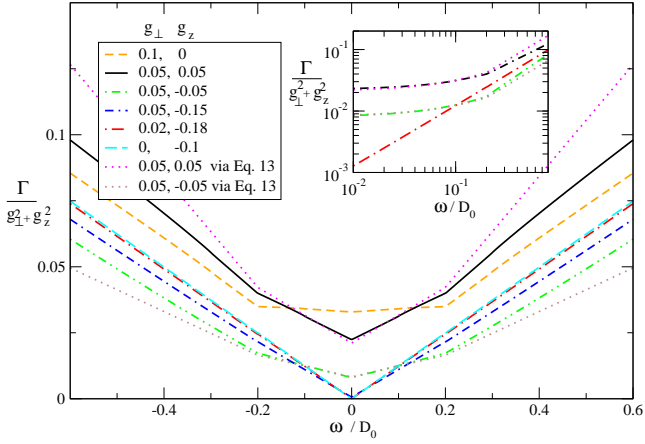}
\end{center}
\par
\vspace{-0.5cm} 
\caption{(Color online) $\Gamma(\protect\omega)$ (rescaled to 
$g_{\perp}^2+g_z^2$ with $g_{\perp,z}$ being the bare Kondo couplings) 
versus $\omega$ at zero 
temperature across the delocalized-localized KT transition. 
 Inset: the 
log-log plot of $\Gamma(\protect\omega)$ versus $\omega$. 
The bias voltage 
is fixed at $V=0.2 D_0$. Here, $D_0 = 1$ for all the figures.}
\label{gamma-w-diss}
\end{figure}
It is worthwhile mentioning 
that unlike the equilibrium RG at finite temperatures 
where RG flows are cutoff by temperature $T$,  
 here in nonequilibrium the RG flows will be cutoff by the 
decoherence rate $\Gamma$, a much lower energy scale than $V$, 
$\Gamma\ll V$. This explains the dips (peaks) structure 
in $g_{\perp (z)}(\omega)$ in Fig.~\ref{gpgzfig}. 
In contrast, the equilibrium RG will lead to approximately frequency 
independent 
couplings, (or ``flat'' functions 
$g_{\perp}(\omega)\approx g_{\perp,z}(\omega=0)$). 
In the absence of field $h=0$, $g_{\perp,(z)}(\omega)$ show 
dips (peaks) at $\omega=\pm V/2$. We use the solutions 
of the frequency-dependent Kondo couplings $g_{\perp, z\sigma}(\omega)$
to compute the dynamical charge susceptibility 
of the resonance-level near the transition.\\

{\bf \it Dynamical decoherence rate and charge susceptibility.} 
We have solved for the dynamical decoherence rate $\Gamma(\omega)$ 
at zero temperature self-consistently 
along with the RG equations Eq.~\ref{RG-Kondo}, and the results are  
shown in Fig.~\ref{gamma-w-diss}. As the general trend,
we find $\Gamma(\omega)$ increases with increasing 
frequency; while it decreases in magnitude at low frequencies 
$|\omega| < V$ as the system 
crossovers from the delocalized to the localized phase. 
is due to  
In additions, in the delocalized phase, 
it shows a singular ``kink-like'' behavior 
at frequencies $\omega=\pm V$, 
separating two different behaviors for $|\omega|>V$ and $|\omega|<V$. 
The curves of $\Gamma(\omega)$ for $|\omega|<V$ 
are closer to the linear behavior than those for $|\omega|>V$. As the system 
moves to the localized phase, $\Gamma(\omega)$ 
for both $|\omega|<V$ and $|\omega|>V$ gradually changes its slopes or  
curvatures and finally merges into a single linear behavior deep in the 
localized phase (see Fig.~\ref{gamma-w-diss}).\\

To understand the above qualitative features, it proves useful 
to simplify the zero temperature dynamical 
decoherence rate in Eq.~\ref{gamma} to the following form:
\begin{eqnarray}
\Gamma(\omega) &=& \frac{3}{4\pi}
[\int_{-V/2-\omega/2}^{V/2+\omega/2}+ \int_{-V/2+\omega/2}^{V/2-\omega/2}]d\epsilon\nonumber \\ 
&\times&g_{\perp}(\epsilon-\omega /2) g_{\perp}(\epsilon+\omega/2)\nonumber \\ 
&+& \frac{3}{4\pi} [\int_{V/2-\omega/2}^{V/2+\omega/2}+\int_{-V/2-\omega/2}^{-V/2+\omega/2}] d\epsilon \nonumber \\
&\times& g_{z}(\epsilon-\omega /2) g_{z}(\epsilon+\omega/2), (|\omega|<V), \nonumber \\
\Gamma(\omega) &=& \frac{3}{4\pi}[\int_{-V/2-\omega/2}^{V/2+\omega/2} + 
\int_{V/2-\omega/2}^{-V/2+\omega/2}]d\epsilon\nonumber \\
&\times& g_{\perp}(\epsilon-\omega /2) g_{\perp}(\epsilon+\omega/2) \nonumber \\
&+& \frac{3}{4\pi} [\int_{V/2-\omega/2}^{V/2+\omega/2} +\int_{-V/2-\omega/2}^{-V/2+\omega/2} ] d\epsilon \nonumber \\
&\times & g_{z}(\epsilon-\omega /2) g_{z}(\epsilon+\omega/2), (|\omega|>V).
\label{gammaT0}
\end{eqnarray}
\begin{figure}[t]
\begin{center}
\includegraphics[width=3cm]{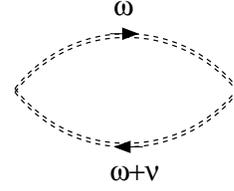}
\end{center}
\par
\vspace{-0.5cm} 
\caption{(Color online) Diagram for the charge susceptibility. 
The dressed pseudofermion propagator (double dashed line) is calculated 
via diagram in Fig.~\ref{self}}
\label{susceptibility}
\end{figure}

In the ``flat'' (or in the ``equilibrium form'') approximation 
where $g_{\perp,z}(\omega)$ are treated as flat functions 
$g_{\perp,z}(V, \omega=0)\approx g_{\perp,z}(V=T)$ 
with $T$ being temperature\cite{chung}, Eq.~\ref{gammaT0} 
can be simplified as the following linear behaviors:
\begin{eqnarray}
\Gamma^e(\omega) &=& \frac{3}{2\pi} 
[V g_{\perp}^2(0) + \omega g_z^2(0)], (|\omega|<V), \nonumber \\
\Gamma^e(\omega) &=&  \frac{3}{2\pi} \omega (g_{\perp}^2(0) +g_z^2(0)), (|\omega|>V)\nonumber \\   
\label{GammaEQ}
\end{eqnarray}
As shown in Fig.~\ref{gamma-w-diss}, the above approximated form 
$\Gamma^e(\omega)$ agrees well with $\Gamma(\omega)$ for $|\omega|< V$, 
but it shows deviation from $\Gamma(\omega)$ for $|\omega|> V$. 
The approximated form $\Gamma^e(\omega)$ exhibits 
two linear behaviors with different slopes for $|\omega|<V$ and 
$|\omega| > V$, respectively, 
leading to a ``kink-like'' singularity at $\omega=\pm V$. 
As the system moves from delocalized to localized phase, 
the ratio $g_{\perp}(0)/g_z(0)$ becomes progressively smaller, leading 
to the suppressions of the two slopes. Finally, as the system is deeply 
 in the localized phase, these two lines 
merge into a single line since $|g_{\perp}(0)| \ll |g_z(0)|$ there. 
The qualitative behaviors of $\Gamma^e(\omega)$ can explain  
the overall monotonically increasing trend of $\Gamma(\omega)$ 
with increasing the magnitude of frequency as well as 
the decreasing trend of $\Gamma(|\omega|<V)$ as the system 
moves from the delocalized to the localized phase.\\ 
\begin{figure}[t]
\begin{center}
\includegraphics[width=8.5cm]{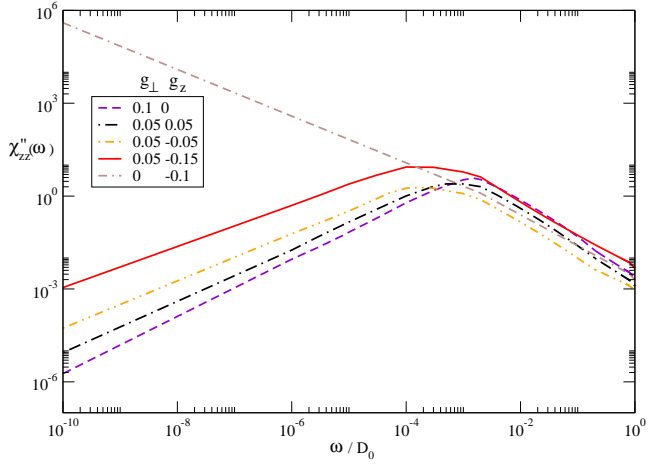}
\end{center}
\par
\vspace{-0.5cm} 
\caption{(Color online) 
$\chi^{"}_{zz}(\protect\omega)$ versus $\omega$ at zero temperature 
across the localized-delocalized KT transition. The 
bias voltage is fixed at $V=0.2 D_0$. Here, $D_0 = 1$ for all the figures.}
\label{chizz_original}
\end{figure}
When the full frequency dependence of $g_{\perp,z}(\omega)$ is considered, 
we find $\Gamma(\omega)$ deviates from the perfect linear behavior in 
$\Gamma^e(\omega)$ (see Fig.~\ref{gamma-w-diss}). Furthermore, 
in the delocalized phase, 
the correction to the linear behavior, $\Gamma(\omega)-\Gamma^e(\omega)$ 
is more noticeable in the large frequency regime ($|\omega|> V$) 
than in the small frequency regime $|\omega|<V$ (see Fig.~\ref{gamma-w-diss}). 
This comes as a result of the wider range in energy $\epsilon$ 
to be integrated over in Eq.~\ref{gammaT0} for $|\omega| > V$, 
which accumulates 
more deviations from the flat approximation due to the dip-peak 
frequency dependence of $g_{\perp,z}(\omega)$. 
As the system moves to the localized phase, $\Gamma(|\omega|> V)$ 
becomes closer to the linear behavior. This is due to the fact that 
$g_z(\omega)$ becomes flatter in the localized phase, giving rise 
to a more linear behavior for $\Gamma(\omega)$. 
Meanwhile, based on the solutions for $g_{\perp,z}(\omega)$, we find 
the correction to the linear behavior 
as well as the singularities at $\omega=\pm V$ are  
logarithmic in nature in the isotropic Kondo limit ($g_{\perp}=0.05=g_z$) 
and at the KT transition ($g_{\perp}=0.05=-g_z$); while 
they are power-law in nature for $|g_{\perp}|\neq |g_z|$ as 
$g_{\perp,z}(\omega)$ show logarithmic and power-law properties in these 
limits, respectively\cite{chung}. 
 Note also that 
for $|\omega| > V$ we find $\Gamma(\omega) < \Gamma^e(\omega)$ 
in the delocalized phase; while the opposite holds in the localized 
phase. This can be understood as in the delocalized phase we have 
$|g_{\perp(z)}(\omega)| < |g_{\perp(z)}(0)|$ 
for the majority of the frequencies (except for $\omega$ very 
close to $\pm V/2$), leading to a smaller value of 
$\Gamma(\omega)$ compared to $\Gamma^e(\omega)$; while the opposite 
is true in the localized phase. We have checked numerically that $\Gamma(\omega=0)$ obtained here indeed 
reproduces the frequency independent decoherence rate $\Gamma$ 
obtained in Ref.~\cite{chung}.\\  
\begin{figure}[t]
\begin{center}
\includegraphics[width=8.5cm]{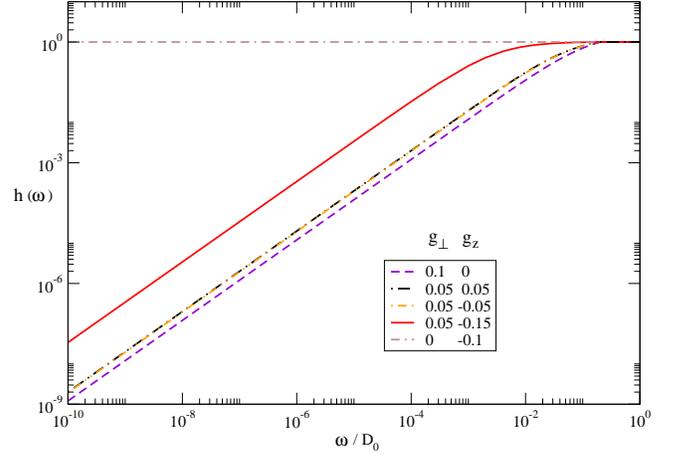}
\end{center}
\par
\vspace{-0.5cm} 
\caption{(Color online) 
$h(\protect\omega)$  
versus $\omega$ at zero temperature 
across the localized-delocalized KT transition. 
The bias voltage is fixed at $V=0.2 D_0$. Here, $D_0 = 1$ for all the figures.}
\label{tanh}
\end{figure}
The effect of the dynamical 
decoherence rate can be measured experimentally via dynamical charge  
susceptibility $\chi_c(\omega)$: 
$Im(\chi_{c}(\omega)) \propto \lim_{h\rightarrow 0} 
d n_d(\omega)/d h$ with $h \propto (N - 1/2)$ being
the effective magnetic field measuring the 
deviations of $N$ electrons on the dot from the charge 
degeneracy point. Here, $\chi_c(\omega)$ is the Fourier-transformed 
charge susceptibility defined as:
\begin{equation}
\chi_c(t) \equiv {\it i} \theta(t) 
<[(n_d(t)-1/2), (n_d(0)-1/2)]>
\end{equation}
Experimentally, the dynamical charge susceptibility 
can be measured by the capacitance lineshape in an AC 
field near the 
charge degeneracy point via the high sensitivity charge sensor 
in the Single Electron Transistor (SET) connected to the dot\cite{berman}. 
The decoherence rate $\Gamma(\omega=0)$ here 
corresponds 
to the broadening of the resonance peak in the imaginary part of 
$\chi_{c}(\omega)$. 
We have calculated $\chi_c(\omega)$ at zero temperature 
based on the renormalized second-order perturbation theory (see the  
diagram in Fig.~\ref{susceptibility}). 
From the mapping mentioned above, 
the dynamical charge susceptibility $\chi_c(\omega)$ is related 
to the $z-$component of the effective ``spin-spin'' correlation 
function $\chi_{zz}$ in the effective Kondo model through:
\begin{eqnarray}
\chi_c(t) &\equiv& {-\it i} \theta(t) 
<[(n_d(t)-1/2), (n_d(0)-1/2)]> \nonumber \\
&\equiv& \chi_{zz}(t),\nonumber \\
\chi_{zz}(t) &\equiv& -{\it i} \theta(t) <[S_z(t), S_z(0)]>
\end{eqnarray}
Taking the Fourier transform of $\chi_{zz}(t)$ and 
evaluating the diagram in Fig.~\ref{susceptibility}, 
the imaginary part of $\chi_{zz}(\omega)$, 
$\chi_{zz}^{"}(\omega)= Im(\chi_{zz}(\omega))$ is given by:
\begin{eqnarray}
\chi_{zz}{"}(\omega) &=& \int \frac{d\epsilon}{2\pi} 
\sum_{\sigma=\uparrow,\downarrow}
[\bar{G}_{f,\sigma}^<(\omega+\epsilon) \bar{G}_{f,\sigma}^>(\epsilon) \nonumber \\
&-& 
\bar{G}_{f,\sigma}^>(\omega+\epsilon) \bar{G}_{f,\sigma}^<(\epsilon)], 
\end{eqnarray}
where $\bar{G}^{>(<))}$ denote the greater (lesser) 
components of the dressed pseudofermion Green's functions: 
\begin{eqnarray}
\bar{G}_{f\sigma}^{<}(\omega)&=& {2\pi \it i} 
n_{f\sigma}(\omega) \bar{A}_{f\sigma}(\omega)\nonumber \\
\bar{G}_{f\sigma}^{>}(\omega)&=& {2\pi \it i} 
(n_{f\sigma}(\omega)-1) \bar{A}_{f\sigma}(\omega)\nonumber \\
\bar{A}_{f\sigma}(\omega) &=& Im[\frac{1}{\omega+ 
\Sigma_{f\sigma}(\omega) +{\it i}\eta}],  
\end{eqnarray}
where $n_{f\sigma}(\omega)$ is the nonequilibrium occupation number determined 
self-consistently by the quantum Boltzmann equation\cite{paaske}: 
\begin{equation}
n_{\sigma}(\omega) = 
(1- \Sigma^{>}_{\sigma}(\omega)/\Sigma^{<}_{\sigma}(\omega))^{-1} 
\end{equation}
where $\Sigma^{>(<)}_{\sigma}(\omega)$ are defined in Ref.~\cite{latha}. 
The resulting expression for $\chi_{zz}(\omega)$ at $T=0$ is given by:
\begin{equation}
\chi_{zz}^{"}(\omega) =  
\frac{\Gamma(\omega)}{\omega^2 + \Gamma^2(\omega)} \times 
(1-2n_{f\uparrow}(\omega))
\label{chizz-w}
\end{equation}
where $n_{f\uparrow}(\omega)$ is given by\cite{paaske}: 
\begin{eqnarray}
n_{f\uparrow}(\omega) &\approx& \frac{1}{2} 
\frac{g_{\perp}^2(0) (V-\omega)}{2g_z^2(0) \omega + g_{\perp}^2(0) (V+\omega)}, 
(0<\omega< V)\nonumber \\
n_{f\uparrow}(\omega) &\approx& \frac{1}{2}
\frac{g_{\perp}^2(0) (V-\omega - 2 g^2_{z} \omega)}{g_{\perp}^2(0) (V+\omega)}, 
(-V<\omega< 0).\nonumber \\
\label{nf}
\end{eqnarray}
Note that we have neglected the vertex correction 
in the calculation for $\chi_{zz}(\omega)$ as it gives only a sub-leading 
correction to Eq.~\ref{chizz-w}. \\

\begin{figure}[t]
\begin{center}
\includegraphics[width=8.5cm]{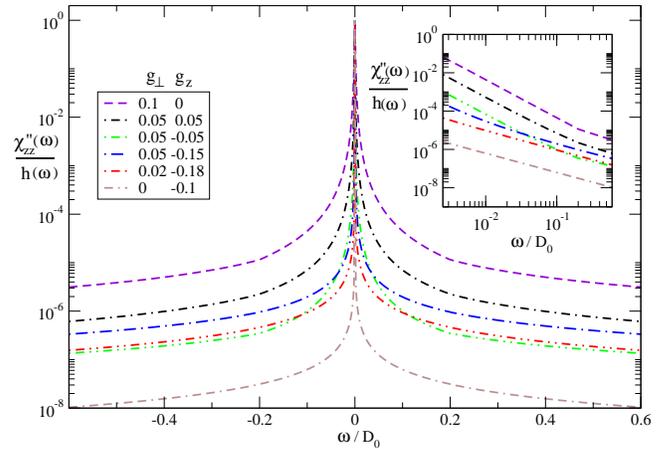}
\end{center}
\par
\vspace{-0.5cm} 
\caption{(Color online) 
$\chi^{"}_{zz}(\protect\omega)/h(\omega)$ (normalized to $1$) 
versus $\omega$ at zero temperature 
across the localized-delocalized KT transition. Here, $h(\omega)$ is defined 
in the text. Inset: the log-log 
plot of $\chi^{'}_{zz}(\protect\omega)$ (normalized to $1$) 
versus $\omega$. The bias voltage is fixed at $V=0.2 D_0$. Here, $D_0 = 1$ for all the figures.}
\label{chizz}
\end{figure}

As shown in Fig.~\ref{chizz}, in the delocalized 
phase, $\chi_{zz}(\omega\rightarrow 0)\propto \omega \rightarrow 0$; while 
$\chi_{zz}(\omega\rightarrow 0)\rightarrow 1/\omega$ in the 
localized phase. 
Hence, as the system crossovers from delocalized 
to localized phase $\chi_{zz}^{"}(\omega)$ shows a dip-to-peak crossover 
for small $\omega$. This behavior can be understood as follows.  
In the delocalized (Kondo screened) phase, the effective local ``spin'' 
get Kondo screened in the low energy scale; therefore, the spin susceptibility 
should vanish. On the other hand, in the localized (ferromagnetic) phase, 
the unscreened free ``spin'' gives rise to the Curie-law susceptibility 
$\chi_{zz}^{"}(\omega)\propto 1/\omega$ at low energies. 
Meanwhile, in the delocalized phase, we find a 
a ``kink-like'' singular behavior  
in $\chi_{zz}^{"}(\omega)$ at $\omega= V$, coming from the 
singular behaviors of both 
$\Gamma(\omega=V)$ (see Eq.~\ref{gamma-w-diss}) 
and the factor $1-2n_{f\uparrow}$ 
in $\chi^{"}_{zz}(\omega=V)$ (see Eq.~\ref{chizz-w} and discussions below). 
However, this singularity gets smeared out as the system crossovers to the 
localized phase. We have checked that our results for 
$\chi_{zz}^{"}(\omega)$ in the isotropic Kondo limit  
agrees qualitatively well with those in Ref.~\cite{schoeller} 
and Ref.~\cite{kehrein2}. For $\omega \gg V$, we find Curie-like 
susceptibility $\chi_{zz}^{"}(\omega)\propto 1/\omega$ in both 
localized and delocalized phases, followed Eq.~\ref{chizz-w} and 
Eq.~\ref{GammaEQ}.

We can furthermore link our results to the 
equilibrium and nonequilibrium fluctuation-dissipation 
theorem\cite{schoeller,coleman,kehrein,kehrein2,mitra}. It is useful to define 
the dynamical fluctuation-dissipation ratio $h(\omega)$:
\begin{eqnarray}
h(\omega) = \frac{ \chi_{zz}^{"}(\omega)}{S_{zz}(\omega)},
\label{h}
\end{eqnarray}
where $S_{zz}(\omega)$ is the Fourier-transformed longitudinal 
spin-spin correlation function with its real-time form given by:
\begin{equation}
S_{zz}(t) = \frac{1}{2}<\{S_z(t), S_z(0)\}>.
\label{Szz}
\end{equation}
The dynamical spin-spin correlation function $S(\omega)$ is given by: 
\begin{eqnarray}
S(\omega) &=& \int \frac{d\epsilon}{2\pi} 
\sum_{\alpha=\uparrow,\downarrow}
[\bar{G}_{f,\alpha}^<(\omega+\epsilon) \bar{G}_{f,\alpha}^>(\epsilon) \nonumber \\
&+& 
\bar{G}_{f,\alpha}^>(\omega+\epsilon) \bar{G}_{f,\alpha}^<(\epsilon)], 
\end{eqnarray}
Carrying out similar calculations as in $\chi_{zz}^{"}(\omega)$, 
we find the fluctuation-dissipation ratio 
$h(\omega)$ reads: 
\begin{equation}
h(\omega) = 1-2 n_{f\uparrow}(\omega),  
\label{h}
\end{equation}
where $n_{f\uparrow}(\omega)$ is defined in Eq.~\ref{nf}.
 In equilibrium, 
the ratio $h(\omega)\equiv h^0(\omega)$ respects the fluctuation-dissipation theorem\cite{coleman,schoeller,kehrein,kehrein2}, given by 
$h^0(\omega) = 1-2 n_f^0(\omega)=\tanh(\beta\omega/2)$ where 
$n_f^0(\omega)= \frac{1}{e^{\beta\omega}+1}$ is the Fermi function for 
pseudofermion in equilibrium. At $T=0$ ($\beta\rightarrow \infty$), we 
have $h^0(\omega>0) =1$, 
the signature of the equilibrium fluctuation-dissipation 
theorem at $T=0$. In nonequilibrium and at $T=0$, however, we find 
in general a deviation of $h(\omega)$ in the delocalized phase 
from the equilibrium fluctuation-dissipation theorem:  
$h(\omega) <1$ for $\omega < V$\cite{schoeller,coleman}, and 
$h(\omega\rightarrow 0)\propto \omega \rightarrow 0$ 
(see Fig.~\ref{tanh} and  
Eq.~\ref{h}, Eq.~\ref{nf}); while we recover the equilibrium 
fluctuation-dissipation theorem $h(\omega)=1$ for $|\omega|> V$, 
in agreement with the results in found in Ref.~\cite{schoeller}. 
However, that as the system moves to the deep 
localized phase, the above deviation gets smaller and finally  
we recover the equilibrium fluctuation-dissipation theorem for 
all frequencies: $h(\omega)=1$ for all $|\omega|<D_0$ (see Fig.~\ref{tanh}). 
To more clearly observe the singular property of 
$\chi^{"}(|\omega|=V)$, we also plot the normalized 
Lorentzian form 
$\Gamma(0)\chi^{"}(\omega)/h(\omega)= \Gamma(\omega)\Gamma(0)/(\omega^2+\Gamma^2(\omega))$ (see Fig.~\ref{chizz}). As the system moves from the delocalized 
to the localized phase, the width of the Lorentzian peak gets 
narrower and the singularity at $\omega=V$ gradually disappears.

%It is worthwhile noting that  
%the good agreement in $g_{\perp,z}(\omega)$ between our FRG approach 
%and the previous results in Ref.~\cite{chung} is expected 
%as it is due to the fact that the 
%difference between $\Gamma(\omega)$ and $\Gamma(\omega=0)$ 
%from our numerical solutions   
%is only sub-leading with respect to the renormalized Kondo coupings. 
%The correction to $g_{\perp,z}(\omega)$ from the frequency 
%dependence of $\Gamma(\omega)$ is small.  
%Nevertheless, the additional singularity in the spectral 
%property of $\Gamma(\omega)$ 
%across the transition comes as a result of the frequency dependence 
%of $\Gamma(\omega)$. This may serve as an 
%additional signature of the localized-delocalized crossover 
%exhibited in $\chi_c^{"}(\omega)$ apart from the change in the  
%width of the peak at $\omega=0$. 

\subsection{\bf Conclusions}

In conclusion, we have calculated the zero temperature 
nonequilibrium dynamical decoherence rate 
and charge susceptibility of a dissipative quantum dot. 
The system corresponds to a nonequilibrium anisotropic Kondo model. 
We generalized previous RG approach based 
on Anderson's poor man's scaling approach and renormalized 
perturbation theory to a more rigorous 
functional RG (FRG) 
approach. Within the FRG approach, as the systems 
crossover from the delocalized to the localized phase, 
we find a crossover in dynamical charge susceptibility from a 
dip to a peak structure for $|\omega| <V$. Meanwhile,  we find 
a smeared-out of the singular behavior in charge susceptibility at 
$\omega=\pm V$ in the above crossover. We also show the deviation of the 
equilibrium fluctuation-dissipation theorem for small frequencies 
$\omega<V$; while the theorem is respected when the system is  
in the extremely localized phase. The above signatures 
can be used to identify experimentally the above localized-delocalized 
quantum phase transition 
out of equilibrium in nano-systems associated with the Kondo effect.

\acknowledgements

We are grateful for the helpful discussions with 
P. W\"{o}elfle. This work is supported by the NSC grant
No.98-2918-I-009-06, No.98-2112-M-009-010-MY3, the MOE-ATU program, the
NCTS of Taiwan, R.O.C..

%\bibitem{Lee}
%Yu-Wen Lee and Yu-Li Lee, Phys. Rev. B {\bf 65}, 155324 (2002).

%\bibitem{Ng}
%Yu-Liang LIu and T. K. Ng, Phys. Rev. B {\bf 61}, 2911 (2000).

%\bibitem{Kim}
%Eugene Kim, cond-mat/0106575 (unpublished).

%\bibitem{wen}
%Xiao-Gang WEn, cond-mat/9812431 (unpublished).


\references

\bibitem{sachdevQPT}
S. Sachdev, {\it Quantum Phase Transitions}, Cambridge University Press (2000).

\bibitem{Steve}
S. L. Sondhi, S. M. Girvin, J. P. Carini, and D. Shahar,
Rev. Mod. Phys. {\bf 69}, 315 (1987).

\bibitem{lehur1}
K. Le Hur, Phys. Rev. Lett. {\bf 92}, 196804 (2004);
M.-R. Li, K. Le Hur, and W. Hofstetter, Phys. Rev. Lett. {\bf 95}, 086406 (2005).

\bibitem{lehur2}
K. Le Hur and M.-R. Li, Phys. Rev. B {\bf 72}, 073305 (2005).

\bibitem{Markus}
P. Cedraschi and M. B\" uttiker, Annals of Physics (NY) {\bf 289}, 1 (2001).

\bibitem{matveev}
A. Furusaki and K. A. Matveev, Phys. Rev. Lett. {\bf88}, 226404 (2002).

\bibitem{zarand}
 L. Borda, G. Zarand, and D. Goldhaber-Gordon, cond-mat/0602019.

\bibitem{Zarand2}
G. Zarand, C.H. Chung, P. Simon, and M. Vojta, 
Phys. Rev. Lett. 97, 166802 (2006).

\bibitem{Josephson}
G. Refael, E. Demler, Y. Oreg, and D. S. Fisher,
Phys. Rev. B {\bf 75}, 014522 (2007).


\bibitem{McKenzie}
J. Gilmore and R. McKenzie, J. Phys. C. {\bf 11}, 2965 (1999).

\bibitem{chung}
C.H. Chung, K. Le Hur, M. Vojta and P. W\" olfle, Phys. Rev. Lett {\bf 102}, 216803 (2009).

\bibitem{chung2}
C.H. Chung, K.V.P. Latha, K. Le Hur, M. Vojta and P. W\"olfle, 
arXiv:1002.1757.

\bibitem{latha}
C.H. Chung and K.V.P. Latha, arXiv:1002.4038 (to appear in Phys. Rev. B).

\bibitem{Rosch}
A. Rosch, J. Paaske, J. Kroha, and P. W\"{o}lfle, 
Phys. Rev. Lett. {\bf 90}, 076804 (2003);
A. Rosch, J. Paaske, J. Kroha, P. W\" offle,
J. Phys. Soc. Jpn. {\bf 74}, 118 (2005).

\bibitem{paaske}
J. Paaske, A. Rosch, J. Kroha, and P. W\"offle, Phys. Rev. B {\bf 70}, 155301 
(2004); J. Paaske, A. Rosch, P. W\"offle, Phys. Rev. B {\bf 69}, 155330 (2004).

\bibitem{woelfleFRG}
H. Schmidt and P. W\"elfle, 
Ann. Phys. (Berlin) {\bf 19}, No. 1-2, 60-74 (2010). 

\bibitem{schoeller}
Severin G. Jakobs, Mikhail Pletyukhov, and Herbert Schoeller, 
Phys. Rev. B {\bf 81}, 195109 (2010). 

\bibitem{kehrein}
Andreas Hackl, David Roosen, Stefan Kehrein, Walter Hofstetter, 
Phys. Rev. Lett. 102, 196601 (2009); P. Fritsch, S. Kehrein, 
Phys. Rev. B 81, 035113 (2010).  

\bibitem{kehrein2}
P. Fritsch and S. Kehrein, Ann. Phys. {\bf 324}, 1105 (2009).

\bibitem{coleman}
W. Mao, P. Coleman, C. Hooley, and D. Langreth, Phys. Rev. Lett. 
{\bf 91}, 207203 (2003).

%\bibitem{shnirman}
%A. Shnirman and Y. Makhlin, Phys. Rev. Lett. {\bf 91}, 207204 (2003).

\bibitem{mitra}
A. Mitra, A. J. Millis, Phys. Rev. B {\bf 72}, 121102 (R) (2005).

\bibitem{berman}
D. Berman, N. B. Zhitenev, R. C. Ashoori, H. I. Smith, and M. R. Melloch, 
J. Vac. Sci. Technol. B {\bf 15}, 2844 (1997); 
D. Berman, N. B. Zhitenev and R. C. Ashoori, and M. Shayegan, Phys. Rev. Lett. 
{\bf 82}, 161 (1999); K. W. Lehnert, B. A. Turek, K. Bladh, L. F. 
Spietz, D. Gunnarsson, P. Delsing, and R. J. Schoelkopf, Phys. Rev. Lett. 
{\bf 91}, 106801 (2003).

\bibitem{Kim}
E. Kim, cond-mat/0106575 (unpublished).

\bibitem{Gogolin}
M. Fabrizio and A. O. Gogolin, Phys. Rev. B \textbf{51}, 17827 (1995).

\bibitem{Florens}
Serge Florens, Pascal Simon, Sabine Andergassen, and Denis Feinberg, 
Phys. Rev. B {\bf 75}, 155321 (2007).
\end{document}